\documentclass[10pt]{article}

\usepackage{amsmath}

\def\<{\langle}
\def\>{\rangle}

\title{Gauge anomalies and the Witten--Seiberg correspondence for
N=1 supersymmetric theories on noncommutative spaces}

\author{
L.~O.~Chekhov${}^{\S}$, A.~K.~Khizhnyakov${}^{\dag}$,\\
\\${}^{\S}$
{\it  Steklov Mathematical Institute,}\\ {\it Gubkin str.8,
Moscow, Russia, 117966}\\ chekhov@mi.ras.ru\\
\\${}^{\dag}$
{\it Physical Department, Moscow State University, }\\ {\it
Moscow, Russia, 119899} \\ bolick@orc.ru}

\date{}

\begin{document}

\maketitle

\begin{abstract}
The explicit form of non-Abelian noncommutative supersymmetric
(SUSY) chiral anomaly is calculated, the Wess--Zumino consistency
condition is verified and the correspondence of the Yang--Mills
sector to the previously obtained results is shown. We
generalize the Seiberg--Witten map to the case of N=1 SUSY
Yang--Mills theory and calculations up to the second order in
the noncommutativity parameter are done.
\end{abstract}

\section{Introduction}
The noncommutative field theories dominate in the modern
theoretical physics. Gauge anomalies have been actively studied
recently both in the case of the Yang--Mills theory on a noncommutative
space \cite{GraciaBondia}, and in SUSY field models~\cite{Grisaru}.
Unfortunately, the results obtained in the SUSY case without imposing
the Wess--Zumino gauge are nonpolynomial in fields and can be
investigated only within a perturbation theory (they are customarily
presented in the form of parametric integrals). This also hinders
the direct verification of the consistency condition~\cite{WessZumino};
however, if we impose the Wess--Zumino gauge, then
these results yield the known expression for the chiral anomaly
in the Yang--Mills sector. In the present paper, we calculate the
expression for the consistent SUSY chiral anomaly in a noncommutative
space using the known expression for the correspondent anomaly in
the commutative space~\cite{Chekhov} obtained using the Pauli--Villars
regularization in the one-loop approximation. Instead of introducing
parametric integrals, we use the $1/m^2$-expansion, where the
regularization parameters~$m$ have the dimension of mass.
The answer is an infinite series in $e_{\star}^{V}-1$ and
$1-e_{\star}^{-V}$,
but admits the verification of the consistency condition. In
the bosonic sector, our answer coincides with the standard expression
for the chiral anomaly in noncommutative space \cite{GraciaBondia}.

We also consider the Seiberg--Witten map between noncommutative and
usual gauge theories (the very term noncommutative means in what follows
a theory on a {\sl noncommutative} space, not just a non-Abelian theory)
\cite{Witten} and construct its generalization to the SUSY case
up to the second order in the noncommutativity parameter~$\Theta$.
Note that because of the chiral projection operators, the vector
component of our answer {\sl does not coincide} with the original
answer by Seiberg and Witten.

A noncommutative superspace is characterized by a Moyal product of
functions, which involves only bosonic coordinates
\cite{Ferrara}:
$$
f(x,\theta)\star
g(x,\theta)\equiv e^{i\partial_{\zeta} \Theta\partial_{\eta}/2}
f(x+\zeta,\theta)g(x+\eta,\theta) \rvert_{\eta=\zeta=0}.
$$
In what follows, we use some properties of this product:
the Leibnitz rule $\partial(f\star g)=\partial f\star
g+f\star\partial g,$ the cyclicity of the product under the integral
sign: $\int f_1
\star f_2 \star f_3=\int f_3 \star f_1 \star f_2$,
the obvious property
$\int f \star g=\int fg$, and the definition of the noncommutative
exponent:
$$
e_{\star}^{f}=\sum_{n=0}^{\infty}\frac{i^n}{n!}f\star\dots\star f.
$$
The notation is by Bagger and Wess \cite{WessBegger}, and the integration
$\int$ implies all the necessary integrations together with the standard
trace operation.

\section{Anomaly}
We consider the massless chiral multiplet of fields
$\Phi(z)=\{\Phi_k\}, \bar{D}\Phi=0$, transformed by a irreducible
representation of a compact gauge group. The fields $\Phi$ interact
with the real vector superfield $V(x)=\{V_{ik}\}$ in a way for the action
to have the standard form
$$
S=\int\,d^8z\bar{\Phi}\star e_{\star}^{V}\star\Phi.
$$
The generating functional
$e^{i\Gamma[V]}=\int\,D\bar{\Phi}D\Phi e^{iS}$ is formally
invariant w.r.t.\ the gauge transformations
$$
e_{\star}^V\to e_{\star}^{-i\bar\Lambda}\star e_{\star}^{V} \star
e_{\star}^{i\Lambda},
$$
where $\Lambda$ is the chiral superfield. Breaking this invariance
results in the anomaly,
$$
\delta_{\Lambda}\Gamma[V]=\mathcal{U}_{\bar{\Lambda}}+
\mathcal{U}_{\Lambda},
$$
which by construction must satisfy the Wess--Zumino consistency
conditions \cite{WessZumino}:
\begin{equation}
\label{WZcc}
  \delta_{\bar{M}}\mathcal{U}_{\bar{\Lambda}}
   - \delta_{\bar{\Lambda}}\mathcal{U}_{\bar{M}}
   = i \mathcal{U}_{[\bar{M},\bar{\Lambda}]_{\star}},
\end{equation}
and the analogous conditions for $\mathcal{U}_{\Lambda}$.

We calculate the anomaly using the following invariant
regularization of $\Gamma[V]:
\Gamma_{reg}=\sum_{i=0}^{\infty}c_i \Gamma_i$, where
\[
   e^{i\Gamma_i}=\int\,\mathcal{D}\bar{\Phi}\mathcal{D}\Phi
   \exp i\int\,\bar{\Phi}\star (e_{\star}^{V}-\frac{m_i^2}{\partial^2})
   \star\Phi,
\]
and the constants $c_i$, $m_i$ satisfy the relations
$\sum c_i=0,\,\sum c_i m_i^2=0,\,c_0=1,\,m_0=0$
(the regularization is removed in the limit $m_i\to\infty$).
After simple transformations, the generating functional becomes
\[
  \Gamma_{reg}[V]=-i\sum_i c_i\sum_{n=2}^{\infty}\frac{1}{n}
   \int\,dz_1\dots dz_n v(z_1)\star \bar{D}^2 D^2 G(z_1,z_2)
   \star\dots\star v(z_n)\star \bar{D}^2 D^2 G(z_n,z_1),
\]
where $v=e_{\star}^{V}-1$ and the Green's function
$$
\bar{D}^2 D^2 G(z_1,z_2)
=i\<\Phi(z_1)\bar{\Phi}(z_2)\>=\frac{\bar{D}^2 D^2}{16(m^2-\partial^2)}
\delta(z_1-z_2).
$$
In the regularized expression, we can use the equations of motion
$D^2(e_\star^V-m^2/\partial^2) \star\Phi=0$ for representing the
expression $\mathcal{U}_{\bar{\Lambda}}$ in the form convenient
for further calculations:
\begin{equation}
\label{start}
  \mathcal{U}_{\bar{\Lambda}}=-16\sum_{i=0}^{\infty}c_i m^2_i
  \sum_{n=1}^{\infty}\int\,dz_0\dots dz_n\bar{\Lambda}(z_0)\star
  G(z_0,z_1)\star v(z_1)\star \bar{D}^2 D^2 G(z_1,z_2)\star\dots\star
  v(z_n)\star \bar{D}^2 D^2 G(z_n,z_0),
\end{equation}
where we have used the identity $D^2\bar{D}^2D^2=16\partial^2D^2$.
Equation~\eqref{start} is our starting point for calculating the anomaly.
Because~$V$ is dimensionless and dimension parameters are absent in
a (massless) theory, no more than four covariant derivatives enter
the anomaly expression. This follows directly from the nonrenormalization
theorem, which claims that any perturbative contribution to an
effective action can be expressed as a single integral over the
superspace. We therefore omit terms with five and more covariant derivatives
when calculating the expression for $\mathcal{U}_{\bar{\Lambda}}$,
because these terms disappear when removing the regularization.
We calculate the anomaly dragging all the covariant derivatives
in~\eqref{start}, except four covariant derivatives, to the left.
We then obtain three cases.

{\bf The first case:} No derivatives act on~$v$. Then, using the
identical transformations
$$
\bar{D}^2D^2\bar{D}^2G_{ij}=\bar{D}^2
(16m^2G_{ij}-\delta_{ij})
$$
we can perform all but one
$\theta$-integrations and obtain after resumming the expression
\[
  \mathcal{U}_{\bar{\Lambda}}^{(1)}=-\sum_{i=0}^{\infty}c_i
  \sum_{n=1}^{\infty} m^{2n}\int\,\bar{\Lambda}(x_0,\theta)\star
  G(x_0,x_1)\star w(x_1,\theta) \star G(x_1,x_2)\star\dots\star
  w(x_n,\theta)\star G(x_n,x_0) d\theta dx_0\dots dx_n,
\]
where
$$
w(x,\theta)=\sum^{\infty}_{n=1}(-1)^{n+1}v_{\star}^{n}=
1-e_\star^{-V(x,\theta)},
$$
and $G(x,y)$ is the standard scalar propagator
$(m^2-\partial^2)^{-1} \delta(x-y)$. In the limit $m\to\infty$,
we obtain
\begin{multline*}
   \mathcal{U}_{\bar{\Lambda}}^{(1)}=\frac{i}{16\pi^2}
   \sum_{n=1}^{\infty} \int\,\sum_{i=1}^{n}\sum_{j=i+1}^{n+1}
   \frac{(j-i)(n+1-j+i)}{n(n+1)(n+2)}\partial_{\mu}^{(i)}\partial^{\mu(j)}
   \bar{\Lambda}^{(n+1)}(x,\theta)\star w^{(1)}(x,\theta)
   \star\dots\star w^{(n)}(x,\theta)
\\ +\frac{i}{16\pi^2}\sum cm^2\ln m^2 \int\,
   \bar{\Lambda}(x,\theta)w(x,\theta),
\end{multline*}
where $\partial^{(i)}$ acts only on the
$i$th line, while $\bar{\Lambda}$ is the
$(n+1)$th line. The second term does not contribute in the anomaly
because it is proportional to the variation of the local (in the
$\star$-product sense as well) functional
$\int e_{\star}^{-V}$ and can be exactly compensated either by
cancellation or by imposing the additional condition
$\sum_i c_im_i^2\ln m_i^2=0.$

{\bf The second case.} Two covariant derivatives act on
$v$-lines. Using the identity $\bar{D}^2D^2
\bar{D}_{\dot{\alpha}}=-4i\bar{D}^2D^{\alpha}\hat{\partial}_
{\alpha\dot{\alpha}}$ and performing the analogous transformations
we then obtain
\begin{multline*}
   \mathcal{U}_{\bar{\Lambda}}^{(2)}=\frac{1}{2}\frac{1}{16\pi^2}
   \sum_{n=2}^{\infty}\int\,\sum_{j=2}^{n}\frac{\sigma^{\mu}_
   {\alpha\dot{\alpha}}}{(n+1)n(n-1)}\left(\sum_{i=1}^{j-1}(n+1+i-j)
   \partial^{(i)}_\mu+\sum_{i=j+1}^{n+1}(i-j)\partial^{(i)}_\mu\right)
\\ \bar{\Lambda}^{(n+1)}\star w^{(1)}\star\dots\star w^{(j-1)}
   \star D^{\alpha}[e_{\star}^V\star\bar{D}^{\dot{\alpha}}e_{\star}^{-V}\star
   w^{(j+1)}\star\dots\star w^{(n)}]
\end{multline*}

{\bf The third case}. All four covariant derivatives act on external
lines. We do not write this case separately. Instead, we add all the
results obtained and write the eventual answer for the chiral anomaly:
\begin{equation}
\label{anomaly}
\begin{split}
   \mathcal{U}_{\bar{\Lambda}}=&\frac{i}{16\pi^2}
   \sum_{n=1}^{\infty}\frac{1}{n(n+1)(n+2)}\int\,\left[
   \sum_{i=1}^{n}\sum_{j=i+1}^{n+1}(j-i)(n+1-j+i)
   \partial^{(i)}\partial^{(j)}\bar{\Lambda}\star w^{n}_{\star}\right.
\\ &-\frac{i}{2}\sum_{j=1}^{n}
   \biggl(\sum_{i=1}^{j}(n+1+i-j)\partial^{(i)}_\mu+
         \sum_{i=j+1}^{n+1}(i-j)\partial^{(i)}_\mu \biggr)
   \bar{\Lambda}\star w^{j}_{\star} \star\sigma^{\mu}_{\alpha\dot{\alpha}}
   D^{\alpha}[e_{\star}^V\star\bar{D}^{\dot{\alpha}}e_{\star}^{-V}\star
   w^{n-j}_{\star}]
\\ &-\frac{1}{16}(n+2)\sum_{i=0}^{n-1}
   \bar{\Lambda}\star w^{n-i}_{\star}\star D^2[(\bar{D}^2e^{V}_{\star})
   \star e^{-V}_{\star}\star w^{i}_{\star}]
\\ &\left.+\frac{1}{4}\sum_{i=0}^{n-1}\sum_{j=0}^{n-1-i}
   \bar{\Lambda}\star w^{n-i-j}_{\star}\star
   D_{\alpha}[e^{V}_{\star}\star\bar{D}_{\dot{\alpha}}e^{-V}_{\star}\star
   w^{i}_{\star}\star D^{\alpha}(e^{V}_{\star}\star\bar{D}^{\dot{\alpha}}
   e^{-V}_{\star}\star w^{j}_{\star})] \right].
\end{split}
\end{equation}

We can now verify consistency conditions~\eqref{WZcc}. The calculations
are quite cumbersome but straightforward and we only give a short
comment. The right-hand side of~\eqref{WZcc} appear when varying
the last variables in each term in~\eqref{anomaly}, while all other
terms are mutually cancelled. We now consider the terms that give
nonzero contribution. Because
$\delta_{\bar{M}}w=-i\bar{M}+iw\star\bar{M}$, we find that
in each term in~\eqref{anomaly}, the inhomogeneous part of the
$w$ variation produces a symmetric over $\bar{\Lambda}$ and $\bar{M}$
expression, which does not contribute to the commutator.
Adding the remaining variations and using the commutation relations
for~$D$ and $\bar{D}$, we obtain
\[
  \delta_{\bar{M}}\mathcal{U}_{\bar{\Lambda}}=
  i\mathcal{U}_{\bar{M}\star\bar{\Lambda}}+(\text{symmetric part under}
  \,\bar{M}\leftrightarrow\bar{\Lambda}),
\]
and, therefore,
$$
\delta_{\bar{M}}\mathcal{U}_{\bar{\Lambda}}
-\delta_{\bar{\Lambda}}\mathcal{U}_{\bar{M}}
=i\mathcal{U}_{[\bar{M},\bar{\Lambda}]_{\star}}.
$$
We were able to verify the consistency conditions only because we have had
the explicit expression for the anomaly. The obtained anomaly expression
could be nonminimal because a part of terms can be variations of local
(in the $\star$-product sense) functionals. However, to simplify
drastically expression~\eqref{anomaly} is difficult because all the terms,
except the first one, are there antisymmetric w.r.t.\ the covariant
derivatives and cannot be reduced to the variations of
$\star$-local functionals.

The comparison of our expression with the answer in \cite{Grisaru}
is difficult as well, because the answer there was expressed as a parametric
integrals of superfields. We therefore calculate only the bosonic part of the
anomaly and show that it coincides with the answer obtained in
\cite{GraciaBondia}. Because noncommutative properties of the space do not
affect the superfield structure \cite{Ferrara}, we can impose the
Wess--Zumino gauge in which the vector component of the superfield~$V$
is $-2\theta\sigma^{\mu}\bar{\theta}A_{\mu}$, while
$V^{3}_{\star}\equiv0$, and all the terms of order five and higher in~$V$
disappear from the anomaly. We introduce the multiplier~$2$ in order to
compare with the standard non-SUSY action
$\bar{\psi}\bar{\sigma}^{\mu}A_{\mu}\psi$. We can now easily calculate
all terms that contribute in the topologically nontrivial part of the
anomaly:
\begin{align*}
  &-\frac{i}{48}\bar{\Lambda}\star(-4\partial V\star D\bar{D}V
  -\partial V\star\bar{D}V\star DV+V\star\bar{D}V\star\partial DV
  -2\partial V\star DV\star\bar{D}V+
\\&\quad\quad +2V\star DV\star\partial\bar{D}V
  +2V\star\partial DV\star\bar{D}V)
\\&-\frac{1}{32}\bar{\Lambda}\star V\star D^2\bar{D}_{\dot{\alpha}}V\star
              \bar{D}^{\dot{\alpha}}V
  +\frac{1}{96}\bar{\Lambda}\star V\star \bar{D}_{\dot{\alpha}}V\star
              D^2\bar{D}^{\dot{\alpha}}V
  -\frac{1}{48}\bar{\Lambda}\star V\star D_{\alpha}\bar{D}_{\dot{\alpha}}V
              \star D^{\alpha}\bar{D}^{\dot{\alpha}}V.
\end{align*}
The direct caclulations using the known properties of the spinor algebra
\cite{WessBegger} then give the answer:
\[
  \mathcal{U}=\mathcal{U}_{\bar{\Lambda}}-\bar{\mathcal{U}}_{\bar{\Lambda}}
  =-\frac{i}{24\pi^2}\int\alpha\star\epsilon^{\lambda\mu\nu\rho}
  \partial_{\lambda}(A_{\mu}\star\partial_{\nu}A_{\rho}+
  \frac{i}{2}A_{\mu}\star A_{\nu}\star A_{\rho}).
\]

\section{The Seiberg--Witten map}
We now turn to constructing the correspondence between superfields
on commutative and noncommutative spaces. We exploit the general
principle by Seiberg and Witten \cite{Witten} that it exists a map
from $V$ to $\hat{V}$ such that
\begin{equation}
\label{WSequation}
  \hat{V}(V)+\hat{\delta}_{\hat{\Lambda}}\hat{V}(V)=
  \hat{V}(V+\delta_\Lambda V).
\end{equation}
We first consider the Abelian case and find the desired map up to the
second order in $\Theta$:
\begin{align*}
  &\hat{V}=V+V_1+V_2+\dots,\quad V_n\sim\Theta^n, \\
  &\hat{\Lambda}=\Lambda+\Lambda_1+\Lambda_2+\dots,\quad
     \Lambda_n\sim\Theta^n.
\end{align*}
Substituting these expressions in \eqref{WSequation} and keeping only terms
up to the second order in $\Theta$, we obtain the equations
\begin{equation}
\label{2equations}
 \begin{split}
  V_1(V+\delta V)-V_1(V)-i(\Lambda_1-\bar{\Lambda}_1)=
     &-\frac{1}{2}\Theta_{ab}(\partial_aV\partial_b\Lambda+
     \partial_aV\partial_b\bar{\Lambda}) \\
  V_2(V+\delta V)-V_2(V)-i(\Lambda_2-\bar{\Lambda}_2)=
     &-\frac{1}{2}\Theta_{ab}( \partial_aV\partial_b\Lambda_1+
     \partial_aV_1\partial_b\Lambda+ \partial_aV\partial_b\bar{\Lambda}_1+
     \partial_aV_1\partial_b\bar{\Lambda}) \\
     &-\frac{i}{12}\Theta_{ab}\Theta_{cd} (\partial_aV \partial_b
     (\partial_cV \partial_d\Lambda)-\partial_aV \partial_b
     (\partial_cV \partial_d\bar{\Lambda})).
 \end{split}
\end{equation}
In order to preserve the chirality of $\Lambda$ we must introduce
the chiral and antichiral projection operators
$P=\bar{D}^2D^2/16\partial^2$ and $\bar{P}=D^2\bar{D}^2/16\partial^2$.
The first equation in \eqref{2equations} then has a unique
solution
\begin{align*}
  &V_1=\frac{i}{2}\Theta_{ab}(1-P)\partial_aV(1-\bar{P})\partial_bV \\
  &\Lambda_1=\frac{i}{2}\Theta_{ab}\partial_a\Lambda P\partial_bV.
\end{align*}
However, because of the projection operators, the obtained expression is
nonlocal and in the Yang--Mills sector becomes
\begin{align*}
  &A_{(1)\mu}=\Theta_{\alpha\beta} (\partial_\alpha A_\mu B_\beta
  -\frac{1}{2}\partial_\mu B_\alpha B_\beta) \\
  &\partial_\mu\lambda_{(1)}=\frac{1}{2}\Theta_{\alpha\beta}(
  \partial_\alpha\partial_\mu\lambda B_\beta),
\end{align*}
where $B_\alpha = P_{\alpha\beta} A_\beta= \partial_\alpha
\partial_\beta/\partial^2A_\beta$ is the longitudinal part of the gauge field
$A_\mu$. However, it is a simple exercise to verify that this formulas {\sl
also} solve the Seiberg--Witten equations for the Yang--Mills fields
$$
\hat{A}_\mu(A)+\hat{\delta}_{\hat{\lambda}}\hat{A}_\mu(A)=
\hat{A}_\mu(A+\delta_\lambda A).
$$

The solution of the second equation in \eqref{2equations} has
already the multiparameteric ambiguity and is rather cumbersome;
we therefore restrict ourselves to the particular solution
\begin{align*}
&V_2=\frac{1}{4} \Bigl[ \Bigr. (1-2P)\partial_aV
    (1-\bar{P})\partial_cV(1-P)\partial_b\partial_dV
    +(1-2\bar{P})\partial_aV(1-P)\partial_cV(1-\bar{P})
     \partial_b\partial_dV \\
&\phantom{V_2=\frac{1}{4}}
    -2\Bigl.(1-P)\partial_aV(1-\bar{P})\partial_cV\partial_b\partial_dV+
    \frac{2}{3}\partial_aV\partial_cV\partial_b\partial_dV
    \Bigr] \Theta_{ab} \Theta_{cd},\\
&\Lambda_2=\frac{1}{2}\partial_a\Lambda P\partial_cV
    P\partial_b\partial_dV\Theta_{ab}  \Theta_{cd}.
\end{align*}
The ambiguity is due to the existence of nonzero solution for the
corresponding equation with the vanishing right-hand side. We can
demonstrate the appearance of such an arbitrariness in solutions
of the equation \eqref{WSequation} to all orders in $\Theta$. Keeping
only terms up to the second order in the gauge superfield $V$,
we obtain the following expressions
\begin{align*}
  &V_1=\frac{i}{2}\Theta_{ab}(1-P)V_a(1-\bar{P})V_b \\
  &V_2=\Theta_{ab}\Theta_{cd} \left(a(1-P)V_{ac}(1-\bar{P})V_{bd}
   +bPV_{ac}PV_{bd}
   +b\bar{P}V_{ac}\bar{P}V_{bd}-\frac{a}{2}V_{ac}V_{bd}\right) \\
  &V_3=-\frac{i}{48}\Theta_{ac}\Theta_{bd}\Theta_{ef}
     (1-P)V_{ace}(1-\bar{P})V_{bdf}
\end{align*}
and analogously in higher orders. Here $V_a=\partial_aV$ and $a,b$
are free parameters. This expression shows that the
ambiguity does not arise in odd orders, and the even orders can be
set to zero by the appropriate choice of the parameters. Summing
up the obtained expressions we get the answer:
\[
  \hat{V}=V+\frac{1}{2}[(1-P)V,(1-\bar{P})V]_\star+O(V^3).
\]

\section{Discussion}
We show that allowing a nonlocality in the Seiberg-Witten map (this
nonlocality necessarily follows from considering $N=1$ SUSY gauge
theories with an unbroken supersymmetry), we obtain a series of
solutions, which do not coincide with the original answer by Seiberg
and Witten. This effect has been shown on the example of Abelian
gauge fields. The interpretation of such an ambiguity as well
as the continuation to a non-Abelian case deserves further
investigations.

\end{document}